\documentclass[pra,preprint,tightenlines,showpacs,twocolumn,10pt]{revtex4} 

\usepackage{epsfig} 
\usepackage{graphicx}
\usepackage{amsmath, amsfonts, amssymb, bm}
\usepackage{braket}
\usepackage{color}
\usepackage[utf8]{inputenc} 
\usepackage{subfigure}

\begin{document}  
    
\title{ Probing atomic 'quantum grating' by collisions with charged projectiles }
 
\author{ S. F. Zhang$^1$, B. Najjari$^1$, X. Ma$^1$ and A. B. Voitkiv$^2$ }
\email{ voitkiv@tp1.uni-duesseldorf.de }
\affiliation{ $^1$ Institute of Modern Physics, Chinese Academy of Sciences, Lanzhou 730000, China \\ 
$^2$ Institute for Theoretical Physics I, Heinrich-Heine-University of D\"usseldorf, Universit\"atsstrasse 1, 40225 D\"usseldorf, Germany } 

\date{\today} 

\begin{abstract}  

The wave function of an atom passed through a  
diffraction grating acquires a regular space structure  
and the interaction of another particle with this atom  
can be thought of as scattering on a 'quantum grating'   
composed of a single atom. 
Probing this 'grating' by collisions with charged projectiles 
reveals interference effects due to coherent contributions 
of its 'slits' to the transition amplitude.  
In particular, the spectra of electrons emitted from the atom 
in collisions with swift ions exhibit 
a pronounced interference pattern whose shape can be    
extremely sensitive to the collision velocity. 

\end{abstract} 


\maketitle 



\section{Introduction} 

One of the fundamental concepts of quantum physics  
is the wave--particle duality: all 
atomic objects exhibit both particle and wave properties. 
It was proposed by Louis de Broglie \cite{dBr92} in 1923. 
Since then a number of experiments confirmed 
the wave nature of various atomic particles ranging 
from electrons \cite{DaG27,Tho28} to  
huge molecules consisting of many thousands of atoms 
\cite{vienna-exper}. 

In most of these experiments atomic particles 
were detected after passing through slits (a diffraction grating). 
In this phenomenon, which is essentially similar 
to the famous Young's double-slits interference of light \cite{Young}, 
the coherent addition of the amplitudes 
for different quantum paths leads to interference in the 
detection probablities clearly demonstrating 
the wave-like behaviour of the particles.   

The diffraction gratings employed in these experiments 
were naturally macroscopic. Their microscopic analogs   
can be found in the atomic world. For instance, 
a diatomic molecule in processes, 
where its internal state does not change, may play 
essentially the same role as the Young's double slits 
in the interaction with light. Since 
interference effects in processes involving 
such molecules arise also when their internal state change, 
it is obvious that -- despite much smaller size -- 
these microscopic 'slits' are associated 
with richer interference effects. Therefore,      
beginning with the work of \cite{Tuan}-\cite{CoF66}, 
very significant efforts have been devoted to 
exploring various interference phenomena arising when particles 
interact with such molecules  
(see e.g. \cite{photo-laser}-\cite{shaofeng-2018}).    

A qualitatively different type of microscopic 'grating'  
can be obtained by letting an atom to pass through a 
macroscopic diffraction grating. 
Indeed, the wave function of the atom  
acquires a periodic space structure. 
As a result, the interaction of another 
particle with this atom can be viewed as scattering 
of the former on an object which can be called a 'quantum grating'. 
Like its macroscopic counterpart 
it possesses a periodic structure, which consists 
of stronger and weaker interacting parts 
corresponding to respectively larger and smaller values 
of the atomic probability density 
but is constructed just of a single atom. 
Thus, the structured probability 
amplitude of a single atom plays now   
essentially the role of a structured macroscopic piece of matter 
in usual gratings.  

In this communication we explore some properties 
of an atomic quantum grating by probing it by the Coulomb 
interaction in collisions with swift charged (structureless) 
projectiles. Interference effects, caused by coherent contributions 
to the transition amplitude from different 'slits' of the quantum grating, 
will be considered for elastic as well as inelastic collisions. 
It, in particular, will be shown that the spectrum of 
electrons emitted in ionizing collisions exhibits pronounced ring-like structures 
whose shape can be extremely sensitive to 
the value of the collision velocity.  

Atomic units are used throughout except where 
otherwise stated. 

\section{ General consideration } 

Let an atom with a momentum $ {\bf P}_a^i = (0, P_a^i, 0)$ pass through  
a macroscopic diffraction grating. The grating is located in the ($x$-$z$)--plane 
and consists of $N_0$ slits along the $z$-direction (see Fig. 1). 
The dimensions of the slits are $ a $ and $ b $ 
(along the $z$- and $x$-directions, respectively), 
the period of the grating is $d$. 

\begin{figure}[t] 
\vspace{-0.25cm}
\centering 
\includegraphics[width=0.49\textwidth]{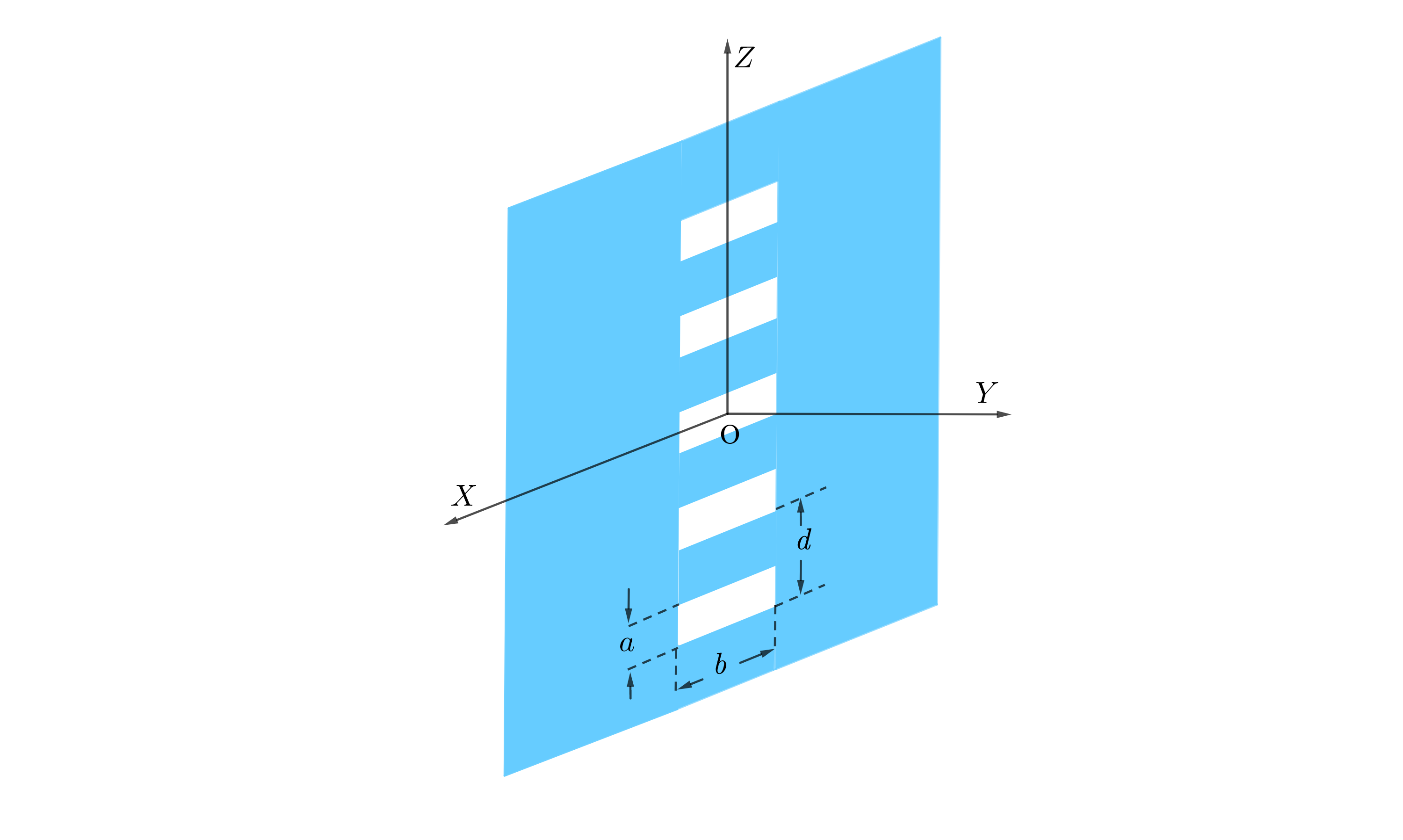}      
\vspace{-0.5cm}
\caption{ \footnotesize{ A diffraction grating with slits located in 
the $x$-$z$ plane, $a$ and $b$ are the slit dimensions and $d$ 
is the period of the grating. } }  
\label{figure1}
\end{figure}  

Before the diffraction grating 
the wavefunction $\psi({\bf R}_a, {\bf r}_e, t)$ 
of the incident atom is given by 
\begin{eqnarray} 
\psi({\bf R}_a, {\bf r}_e, t) &=& \psi_{in}({\bf R}_a, {\bf r}_e, t) 
\nonumber \\ 
& = & A \, \exp(i \, ( {\bf P}_a^i \cdot {\bf R}_a - E_a^i t)) \, 
\phi_i({\bf r}_e-{\bf R}_N),   
\label{incident_atom}
\end{eqnarray} 
where $A$ is the normalization constant, 
${\bf R}_a = (X_a, Y_a, Z_a)$, where $ X_a =    
R_a \sin \vartheta_{{\bf R}_a} \cos \varphi_{{\bf R}_a}$, 
$Y_a = R_a \sin \vartheta_{{\bf R}_a} 
\sin \varphi_{{\bf R}_a}$ and $Z_a = R_a \cos \vartheta_{{\bf R}_a}$,   
is the coordinate of the atomic center-of-mass, 
${\bf r}_e$ and $ {\bf R}_N $ 
are the position vectors of the atomic electron and the atomic nucleus, respectively; 
the origin of the coordinate system is placed 
in the middle of the diffraction grating (see Fig. 1). 
Further, $\phi_i$ is the initial atomic internal state with 
an energy $\varepsilon_i$ and  
$ E_a^i = ({\bf P}_a^i)^2/(2 M_A) + \varepsilon_i $ is the  
total initial energy of the atom with $M_A$ being the atomic mass. 

Using the Huygens-Fresnel principle we obtain 
that after the passage through the grating the wave function 
of the atom at asymptotically large distances ($R_a \gg 
\max \{ N_0 \, d, \, b \} )$  
can be approximated as  
\begin{eqnarray} 
\psi({\bf R}_a, {\bf r}_e, t) &=& \psi_{diff}({\bf R}_a, {\bf r}_e, t)  
\nonumber \\ 
&=& \frac{4 A}{ R_a } \exp(i (P_a^i \, R_a - E_a^i  t)) \,   
\nonumber \\ 
&& \times \frac{ \sin \left( \frac{ P_a^i \, a \, Z_a}{ 2 \, R_a} \right) }
{ P_a^i \, \frac{ Z_a }{ R_a} } \, \, \, \frac{ \sin \left( \frac{ P_a^i \, b \, X_a}{ 2 \, R_a} \right) }
{ P_a^i \, \frac{ X_a }{ R_a} } 
\nonumber \\  
&& \, \times \frac{ \sin \left( N_0 \frac{ P_a^i \, d \, Z_a}{ 2 \, R_a} \right) }{ 
\sin \left( \frac{ P_a^i \, d \, Z_a }{ 2 \, R_a }  \right) } \, 
\phi_i({\bf r}_e - {\bf R}_N). 
\label{diffracted_atom_0}
\end{eqnarray} 
We note that the application of  
the Huygens-Fresnel principle to describe (optical) diffraction  
is known to yield excellent results provided the wavelength 
of light is smaller than $ \min \{ a, b  \} $. In what follows we assume that 
$P_a^i \simeq 10-100$ a.u. and the wavelength of the atom, 
$\lambda_a^i = 1/P_a^i $, will certainly be much smaller than $ \min \{ a, b  \} $  
which enables one to expect that (\ref{diffracted_atom_0}) represents 
a good approximation for the wave function at asymptotically large 
distances between the atom and the grating ($ R_a \gg \max \{ N_0 \, d, \, b \} $).   

As it follows from (\ref{diffracted_atom_0}) the wave function 
of the atom acquired -- due to diffraction -- 
a regular space structure. 
This structure can be unveiled in many ways. 
In what follows we consider how it can be probed 
by letting the atom to interact with a projectile 
which we take here as a point-like charged particle 
(an electron or a bare nucleus).  
 
Assuming that the energy of the projectile is sufficiently high, 
we approximate its initial and final states by plane waves  
\begin{eqnarray} 
\Phi_i({\bf R}_p, t) &=& \frac{ 1 }{ \sqrt{V_p} } \, \exp(i \, ( {\bf p}_p^i \cdot {\bf R}_p \, - \, \varepsilon_p^i \, t)) 
\nonumber \\ 
\Phi_f({\bf R}_p, t) &=& \frac{ 1 }{ \sqrt{V_p} } \, \exp(i \, ( {\bf p}_p^f \cdot {\bf R}_p \, - \, \varepsilon_p^f  \, t)).  
\label{proj_in_and_fin}
\end{eqnarray}   
Here ${\bf R}_p$, ${\bf p}_p^i$ and 
$\varepsilon_p^i = {{\bf p}_p^i}^2/(2 M_p)$ 
(${\bf p}_p^f$ and $\varepsilon_p^f = {{\bf p}_p^f}^2/(2 M_p)$) are the position vector, 
the initial (final) momentum and energy, respectively, of the projectile, 
$M_p$ is its mass and $V_p$ is 
the normalization volume. 
 
Assuming that the momentum of the atom after the collision 
is detected, we take its final state as a product of a plane wave,  
which describes the motion of its center-of-mass, 
and a final internal state of the atom $ \phi_f $ with an energy 
$\varepsilon_f$,     
\begin{eqnarray} 
\psi_f({\bf R}_a, {\bf r}_e, t) = \frac{ \exp(i \, ( {\bf P}_a^f \cdot {\bf R}_a  - E_a^f t)) }{ \sqrt{V_a} }  \phi_f({\bf r}_e - {\bf R}_N), 
\label{atom_fin}
\end{eqnarray}   
where $ {\bf P}_a^f $ and $ E_a^f = ({\bf P}_a^f)^2/(2 \, M_A) + \varepsilon_f$ 
are the final momentum and energy of the atom, respectively, and 
$V_a$ is the normalization volume.  
The state $ \phi_f $ can be either a bound or continuum state; 
the latter case corresponds to ionization. 
 
In the first order of perturbation theory 
the transition amplitude reads 
\begin{eqnarray} 
S_{fi} = - i \int^{+\infty}_{-\infty} dt \, \,  
\langle \Psi_f(t) \mid \hat{W} \mid \Psi_i(t) \rangle.   
\label{trans_ampl}
\end{eqnarray}   
Here $\Psi_i(t) = \psi_{diff}({\bf R}_a, {\bf r}_e, t) \, \Phi_i({\bf R}_p, t) $ 
and $\Psi_f(t) = \psi_f({\bf R}_a, {\bf r}_e, t) \, \Phi_f({\bf R}_p, t) $ are the initial 
and final states, respectively, of the noninteracting  
"atom + projectile" system and 
$ \hat{W} = - Z_p/| {\bf r}_e - {\bf R}_p | + Z_p Z_N/| {\bf R}_N - {\bf R}_p | $, 
where $Z_p$ ($Z_N$) is the charge of the projectile (the atomic nucleus),  
is the interaction between them. 

Let the target and projectile beams cross 
at a distance $D$ from the grating. At this distance  
a single target atom is localized (along the $z$-axis) 
within a spot whose size $\Delta_z$ can be estimated from 
the form of the state (\ref{diffracted_atom_0}) and  
is roughly given by 
$\Delta_z \simeq D/( P_a^i \, a ) \sim  D \, \lambda_a^i/ a  $. 
Taking $ D \simeq 200 $ - $1000$ mm, $a \simeq 10^{-3}$ - $0.1$ mm and 
$P_a^i \simeq 10$ - $100$ a.u. we obtain $\Delta_z \sim 10^{-7}$ - $10^{-3}$ mm. 
This value is much smaller than 
the size $ L_z = N_0 \, d $ ($ \simeq 10^{-2}$ - $1$ mm)  
of the macroscopic grating along the $z$-axis. 
Comparing $ \Delta_z $ and $ L_z $ we see that the typical size 
of the target spot, which produced by all target atoms passed through 
the grating, is essentially determined by $N_0 \, d \sim 10^{-2}$ - $1$ mm and 
thus is much smaller than the distance $D$.  

A similar consideration shows that for a typical size 
of the target spot $\Delta_x$ along the $x$-axis one obtains 
$\Delta_x \simeq \frac{a}{b} \, \Delta_z$ and, assuming 
$b \simeq a$, one has $ \Delta_x \sim 10^{-7}$ - $10^{-3}$ mm. 

Using the above estimates and also noting 
that the diameter of the projectile beam in 
collisional experiments is typically of the order of $1$ mm  
(see e.g. \cite{ref-on-beam-size}) 
we obtain that each dimension of the interaction volume 
(the space volume where the beams cross) 
is much smaller than the distance $D$ between 
its center and the macroscopic diffraction grating. 

Taking all this into account it is convenient 
to introduce a new coordinate system with 
the origin placed in the center of 
the interaction volume. 
The new (primed) and old (unprimed) coordinates 
of the particles are related by 
$ {\bf R}_a' = {\bf R}_a - {\bf D} $,     
$ {\bf R}_p' = {\bf R}_p - {\bf D} $, 
$ {\bf r}_e' = {\bf r}_e - {\bf D} $, 
$ {\bf R}_N' = {\bf R}_N - {\bf D} $, 
where $ {\bf D} =(0, D, 0)$.   
The state of the diffracted atom now reads 
\begin{eqnarray} 
\psi_{diff}({\bf R}'_a, t) & \approx &   
4 \, A \,  \frac{ \exp(i \, P_a^i \, D ) }{ D } 
\exp(i \, (P_a \, Y' \, - \, E_a^i \, t)) \, \, \,  
\nonumber \\ 
& \times & 
\frac{ \sin \left( \frac{ P_a^i \, b }{ 2 } \frac{ X' }{ D } \right) }
{ P_a^i \, \frac{ X' }{ D } }
\, \, \, \frac{ \sin \left( \frac{ P_a^i \, a }{ 2 } \frac{ Z' }{ D } \right) }
{ P_a^i \, \frac{ Z' }{ D } } 
\nonumber \\ 
& \times & \frac{ \sin \left( N_0 \frac{ P_a^i \, d }{ 2 } \frac{ Z' }{ D } \right) }{ 
\sin \left( \frac{ P_a^i \, d }{ 2 } \frac{ Z' }{ D }  \right) } \, 
\phi_i({\bf r}_e' - {\bf R}_N'). 
\label{diffracted_atom}
\end{eqnarray} 
The form of the states (\ref{proj_in_and_fin})-(\ref{atom_fin})  
in the new coordinates are obvious. 

The evaluation of the integrals over 
the interaction volume and time in (\ref{trans_ampl}) results in 
\begin{eqnarray} 
S_{fi} &=& 2^6 \, \pi^3 \, A \, i \, \delta( (E_a^f + \varepsilon_p^f) - (E_a^i + \varepsilon_p^i) )
\nonumber \\ 
& \times & \frac{ \langle \phi_f( \bm \xi ) \mid \exp(i {\bf q} \cdot {\bm \xi}) 
\mid \phi_i( {\bm \xi} )  \rangle }{  q^2  } \,  F_b( q_x  - P_{a, x}^f) 
\nonumber \\ 
& \times & \frac{ \exp(i \, P_a^i \, D ) }{ D } \, 
\times \, \delta(P_a^i + q_y  - P_{a, y}^f )  
\nonumber \\ 
& \times & \sum_n F_a( q_z  - P_{a, z}^f - P_a^i \, d \, n/D ), 
\label{trans_ampl_1}
\end{eqnarray}   
where $\bm \xi = {\bf r}_e' - {\bf R}_N'$, 
$ {\bf q} = (q_x, q_y, q_z) = {\bf p}_p^i - {\bf p}_p^f = 
(p_{p, x}^i - p_{p, x}^f, p_{p, y}^i - p_{p, y}^f,  p_{p, z}^i - p_{p, z}^f)$ 
is the change in the projectile momentum 
($=$ the momentum transfer to the atom) and 
\begin{equation} 
F_\alpha(\eta) =  \left \{ 
\begin{array}{rl}  
& \pi; \, \, | \eta | <  \frac{ P_a^i \, \alpha }{ 2 \, D } \\ 
\\ 
& \pi/2; \, \, | \eta |= \frac{P_a^i \, \alpha }{ 2 \, D } \\ 
\\ 
& 0;  \, \, | \eta | > \frac{P_a^i \, \alpha }{ 2 \, D },  
\end{array}  
\right.
\label{momentum-function} 
\end{equation}
where $\alpha = a, b$. 
It follows from (\ref{trans_ampl_1})-(\ref{momentum-function}) 
that along the $x$- and $z$-directions there are uncertainties,  
$ \Delta P_x = P_a^i \, b/D $ and $ \Delta P_z = P_a^i \, a/D $,  
in the momentum balance in the collision. This is caused by the 
(uncontrolled) momentum exchange 
between the atom and the macroscopic diffraction grating resulting in 
an uncertainty in the momentum of the atom passed through it. 
However, the energy balance in the collision does not possess 
an uncertainty because the macroscopic grating, which has 
essentially infinite (on the atomic scale) mass and is initially at rest, 
does not participate in the energy exchange.     

\section{ Results and discussion } 

\subsection{ Elastic collisions with electrons } 
  
\begin{figure}[t] 
\vspace{-0.25cm}
\centering
\begin{center} 
\includegraphics[width=0.51\textwidth]{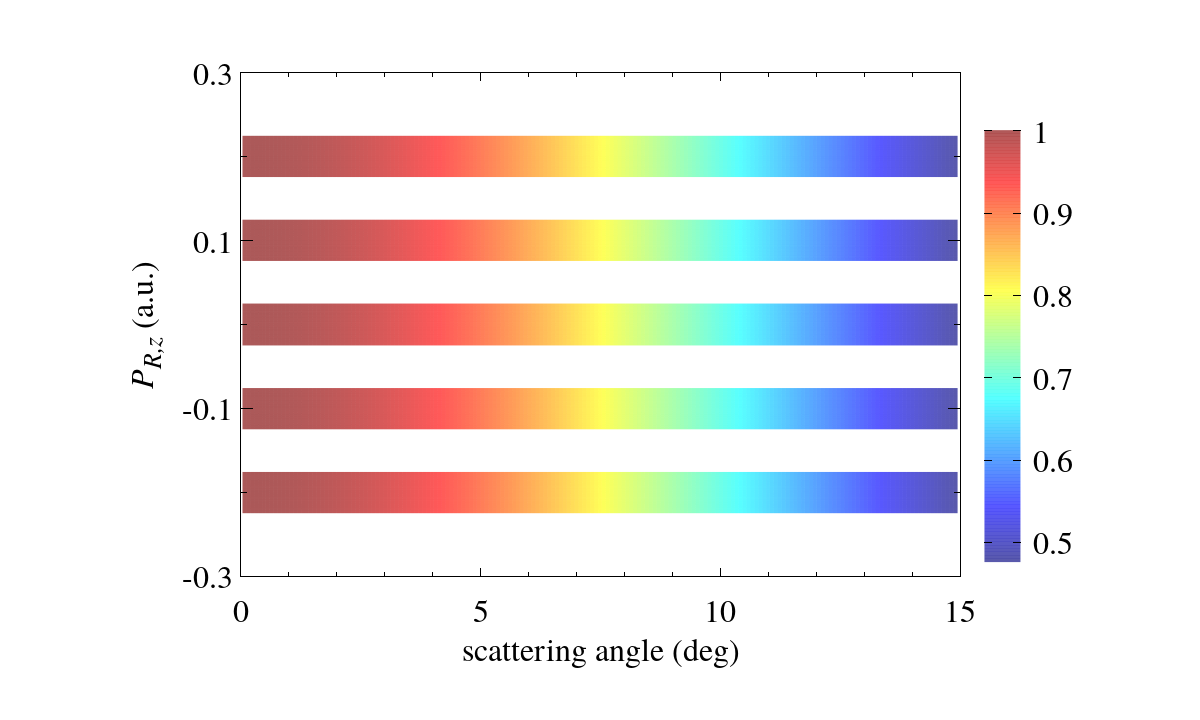}      
\end{center}
\vspace{-1.1cm} 
\caption{ \footnotesize{ The differential cross section for 
elastic scattering of $300$ eV electrons incident along 
the $x$-axis on Xe atoms ($P_a^i = 100$ a.u.) passed through 
a diffraction grating with $a = b = d/2 = 0.1$ mm 
and $N_0 = 5$. $D = 200$ mm. The cross section is given 
in relative units with its intensity ranging from $0$ to $1$. } }  
\label{figure3}
\end{figure}  
Fig. \ref{figure3} displays the elastic cross section, 
differential in the projectile scattering angle and 
the component $P_{a,z}^f$ of the final atomic momentum,  
for $300$ eV electrons incident along the $x$-axis 
on Xe atoms which passed through a diffraction grating.  
A clear interference pattern observed in the cross section 
is caused by the coherent addition of the contributions of 
the different 'slits' of the atomic quantum grating to 
the process of electron scattering.   
 
\subsection{ Ionization in collisions with bare ions } 
  
\begin{figure}[t] 
\vspace{-0.25cm}
\centering
\begin{center} 
\includegraphics[width=0.51\textwidth]{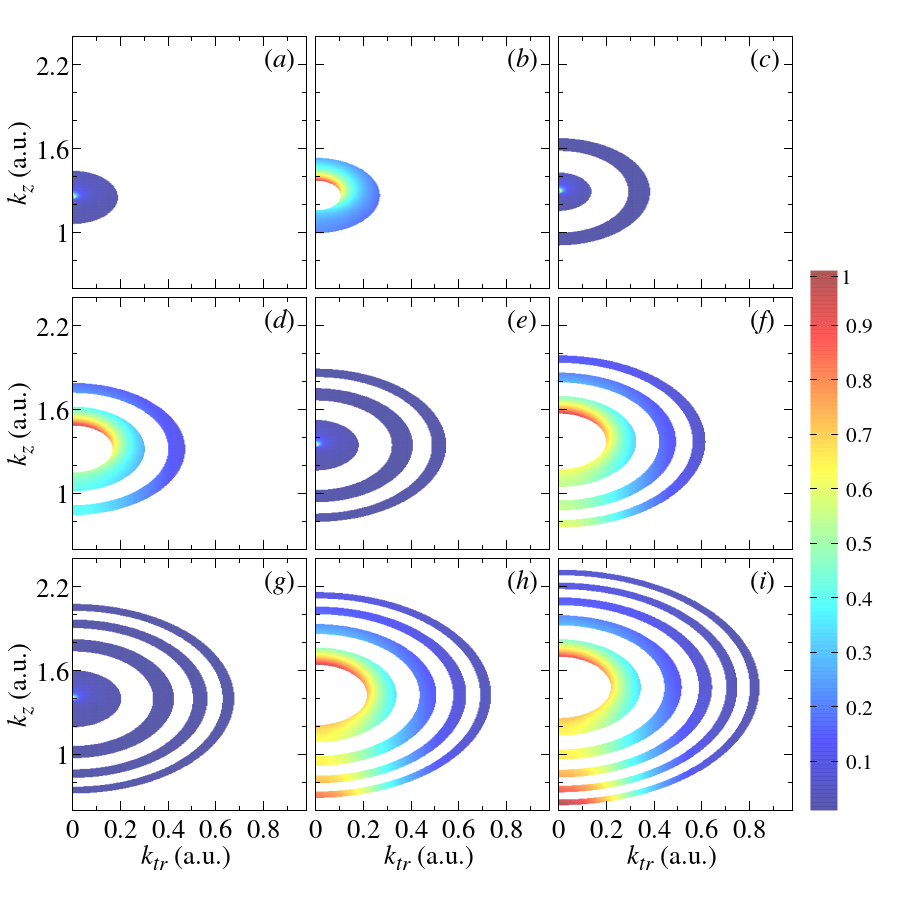}      
\end{center}
\vspace{-0.15cm}
\caption{ \footnotesize{ Spectra of electrons emitted from He atoms 
($ P_a^i = 50 $ a.u., $\varepsilon_i = -0.9$ a.u.) passed 
through a diffraction grating ($a = b = d/2 = 0.1$ mm, 
$ N_0 = 5 $) in collisions with protons. $D = 200 $ mm,  
$ P_{R, z} = q_x = q_y = 0$. 
The collision velocity $v = f \, v_0$, 
where $v_0 = \sqrt{ 2 |\varepsilon_i|  }$ and 
$f = 0.93$ (a), $0.94$ (b), $0.96$ (c), $0.98$ (d), 
$1$ (e) $1.02$ (f), $1.04$ (g) $1.06$ (h), $1.1$ (i). 
The spectra are given in relative units with 
their intensities ranging from $0$ to $1$. } }  
\label{figure4}
\end{figure}  
We now consider collisions of atoms with bare ions  
resulting in atomic ionization. 
Using the transition amplitude (\ref{trans_ampl_1})  
we obtain the fully differential cross section section 
for atomic ionization 
\begin{eqnarray} 
\frac{ d \sigma }{ d^3 {\bf q} \, \, d^3 {\bf P}_R  \, \, d^3 {\bf k} } 
& = & 2^{12} \pi^4 \, \, A^2   \, \frac{ Z_p^2 }{ v \, { P_a^i}^2 }   
\nonumber \\  
& \times & \frac{ \left| \langle \phi_{\bf k}( \bm \xi ) 
\mid \exp(i {\bf q} \cdot {\bm \xi}) 
\mid \phi_i( {\bm \xi} )  \rangle \right|^2 }{  q^4  }  
\nonumber \\ \,    
& \times & \delta(P_a^i + q_y - P_{R, y} - k_y ) \,    
\nonumber \\ 
& \times & \delta(   E_a^f + \varepsilon_p^f - E_a^i - \varepsilon_p^i ) 
\nonumber \\ 
& \times & F_b^2( q_x - P_{R, x} - k_x )   
\nonumber \\ 
& \times & \sum_n F_a^2(B_n). 
\label{fdcs-1}
\end{eqnarray}
Here, ${\bf P}_R = (P_{R, x}, P_{R, y}, P_{R, z})$ is the momentum of the target recoil ion, 
${\bf k} = ( k_x, k_y, k_z )$ is the momentum of the emitted electron, and 
$ B_n = q_z - P_{R, z} - k_z - P_a^i \, d \, n/D$. 

Since $M_A \gg m_e$ ($m_e$ is the electron mass)  
one has $ E_a^f - E_a^i \approx k^2/2 - \varepsilon_i $. 
Besides, for collisions with $|{\bf q}| \ll |{\bf p}_i|$, one obtains 
$\varepsilon_p^i - \varepsilon_p^f = ({\bf p}^2_i - {\bf p}^2_f)/(2 M_p) = 
({\bf p}_i - {\bf p}_f) ({\bf p}_i + {\bf p}_f)/(2 M_p) \approx {\bf q} \cdot {\bf v} $. 
Therefore, the energy delta-function 
in (\ref{fdcs-1}) can be rewritten as  
$ \frac{ 1 }{ v } \delta( q_{\parallel}   - (k^2/2 - \varepsilon_i)/v )$, 
where $ q_{\parallel} = {\bf q} \cdot {\bf v}/v $. 

Let the projectile be incident along the $z$-direction. 
Integrating the cross section (\ref{fdcs-1}) over 
$P_{R, x}$, $P_{R, y}$ and $q_z$ we obtain 
\begin{eqnarray} 
\frac{ d \sigma }{ d^2 {\bf q}_{\perp} d P_{R, z} d^3 {\bf k} } 
& \sim & \frac{ Z_p^2 }{ v^2} \, 
\frac{ \left| \langle \phi_{\bf k}( \bm \xi ) 
\mid \exp(i {\bf q}_0 \cdot {\bm \xi}) 
\mid \phi_i( {\bm \xi} )  \rangle \right|^2 }{  q_0^4  }   
\nonumber \\     
& \times & \sum_n F_a^2(B_n),   
\label{fdcs-2}
\end{eqnarray}
where ${\bf q}_0 = (q_x, q_y, q_{min}) = ({\bf q}_{\perp}, q_{min})$ with 
$q_{min} = (k^2/2 - \varepsilon_i)/v$ and  
$B_n = q_{min} - P_{R, z} - k_z -  P_a^i \, d \, n/D $. 
 
The binary-encounter emission and electron capture 
to the projectile continuum 
belong to the most prominent features of ionization 
(see e.g. \cite{be+ecc} and references therein).  
The former is a two-body ionization mechanism in which 
the momentum exchange 
in the collision occurs between the projectile and the electron 
while the target nucleus/core is merely a spectator: 
$P_{R, x} \approx 0$, $ P_{R, z} \approx 0$, $P_{R, y} \approx P_a^i$.  
In the latter the emitted electron moves with a velocity ${\bf v}_e \approx {\bf v}$, i.e. 
together with the projectile but without forming a bound state with it. 

By analysing the collision momentum balance $q_{min} \approx P_{R, z} + k_z$ 
one can convince oneself that, provided $|\varepsilon_i| = v^2/2$, 
these two features are present simultaneously. 
Such an interesting situation 
is considered in Fig. \ref{figure4} where the cross section  
$ d \sigma/(d^2{\bf q}_{\perp} d P_{R, z} dk_z dk_{\perp}) $, 
taken at $ P_{R,z} = q_x = q_y = 0 $,    
is shown as a function of $k_z$ and $k_{\perp}= \sqrt{ k_x^2 + k_y^2 }$  
for single ionization of atomic helium by proton projectiles. 

As is well known \cite{be+ecc}, for electron capture 
to the projectile continuum the interaction between 
the projectile and the emitted electron is crucial. 
Since the first order approximation neglects this interaction, 
the cross section was calculated using 
the Continuum-Distorted-Wave-Eikonal-Initial-State approach  
\cite{cdw-eis} which accounts for this interaction 
yielding a satisfactory description of this process. 

The electron spectrum shown in Fig. \ref{figure4}  
exhibits a pronounced interference structure consisting of concentric rings. 
The center, the width and the number of the rings 
are determined by the inequalities 
$ B^- \leq (k_z - v)^2 + k_{\perp}^2 \leq B^+ $, where 
$B^{\pm} =  v^2 - 2 |\varepsilon_i| + 2 \, v \, P_a^i \, \frac{ d }{ D } \, n  \, \pm \,  
v \, P_a^i \, \frac{ a }{ D }  $. 
This structure arises due to the coherent contributions to the emission 
from different parts of the atomic quantum grating 
which strongly interfere in the cross section. 
The number of the rings, their size and even the relative intensity on them
turn out to be extremely sensitive to the magnitude of the collision velocity.     

\section{ Conclusions } 

Due to a regular space structure 
acquired by the wave function of an atom, which passed through 
a macroscopic diffraction grating, this atom can itself be regarded  
as a diffraction grating. Compared to 
macroscopic gratings and even to microscopic gratings 
represented by molecules, this type  
is qualitatively different since it consists of just a single atom;  
the role of the structured "real matter" in the former ones 
is now mimicked by the structured probability amplitude.  

We have considered interference effects appearing  
when such an atomic quantum grating is probed by collisions 
with point-like charged projectiles. These effects arise due 
to the coherent addition of the contributions of 
different 'slits' of the atomic grating 
to the transition amplitude and very clearly manisfest themselves 
both in elastic and ionizing collisions.     

In particular, interesting patterns arise in 
the spectra of electrons ejected in collisions with ions 
in the kinematic range where there is a 'confluence' 
of the binary-encounter emission 
and electron capture to the projectile continuum. 
In such a case a pronounced interference structure appears  
in the emission spectrum whose shape is extremely sensitive 
to the magnitude of the collision velocity.  
This point might potentially be exploited 
for a very precise determination of this velocity.  

As preliminary estimates show, atomic quantum gratings  
can profoundly manisfest themself also in 
other basic atomic collision processes (e.g. 
electron capture to a bound state 
and projectile-electron loss) as well as in 
photoabsorption and photon scattering. 

An experimental verification of the theoretical predictions 
requires both (i) high quality (monochromatic) beams of projectiles 
and target atoms and (ii) an accurate determination of the momenta 
of the particles in the final state. 
While there is no big problem with fulfilling the condition (i) 
(for instance, the typical energy spread of $10 - 100$ keV/u ions 
can be optimized to a few eV meeting the monochromatic beam criteria), 
the determination of the final momenta with 
the necessary accuracy is quite challenging.  

In experiments with reaction microscopes helium targets have momenta 
susbtantially less than the value $P_i = 50 $ a.u. used here 
(see Figs. \ref{figure3} and \ref{figure4}). Smaller $P_i$ reduce 
the size and thickness of the rings in Fig. \ref{figure4} setting 
even a stronger requirement on the accuracy in the momentum determination. 
To prepare the initial target moment of $P_i \sim 50 - 100$ a.u., 
one could use heavier targets (e.g. Ar or Xe). But for heavy targets 
the accuracy of the momenta determination is usually of the order of 
a few atomic units for a reaction microscope equipped with 
the room-temperature super-sonic gas jet, which is far away from 
the value of $ \lesssim 0.1$ a.u. necessary to resolve 
the structures in Figs. \ref{figure3} and \ref{figure4}. 
To reach a much better accuracy a reaction microscope 
with a precooling super-sonic gas jet can be envisaged.  

However, the requirements on the accuracy in the determination 
of the final momenta can be greatly softened if, 
instead of target atoms, structured projectiles 
(e.g. He$^+$) pass through a macroscopic grating. 
Due to much higher incident momenta, the uncertainty 
in the projectile momentum after the diffraction grating 
will be much larger. For instance, 
for $25$ - $100$ KeV/u He$^+$ ions passed through a grating with 
$a \simeq b \simeq 0.1$ mm, one obtains 
$\Delta p_p \simeq p_p^i \, a/D \simeq 3.5 - 7$ a.u. at $ D = 200 $ mm.  

\section{ Acknowledgement } 
 
We acknowledge the support from the National Key Research 
and Development Program of China (Grant No. 2017YFA0402300), 
the CAS President's International Fellowship Initiative  
and the German Research Foundation (DFG) under Grant No 349581371 
(the project VO 1278/4-1).  
A. B. V. is grateful for 
the hospitality of the Institute of Modern Physics.


\begin{thebibliography}{99} 

\bibitem{dBr92} L.~de Broglie, PhD thesis, 
reprinted in Ann. Found. Louis de Broglie 17, 22 (1992). 

\bibitem{DaG27} C.~J.~Davisson and L.~H.~Germer,
                Phys. Rev. 30, 705 (1927).

\bibitem{Tho28} G.~P.~Thomson, 
                Proceedings of the Royal Society (London) A117, 600 (1928). 

\bibitem{vienna-exper} Y. Y. Fein, F. Geyer, P. Zwick, F. Kialka, S. Pedalino, M. Mayor, 
S. Gerlich and M. Arndt,  
Nature Physics, https://www.nature.com/articles/s41567-019-0663-9 (2019). 

\bibitem{Young} T. Young, Philosophical Transactions, Nov. 24th 1803.  
 
\bibitem{Tuan} T. F. Tuan and E. Gerjuoy, 
Phys. Rev.  {\bf 117}, 756 (1960).  

\bibitem{CoF66} H.~Cohen and U.~Fano, Phys. Rev. 150, 30 (1966).


\bibitem{photo-laser} F. Lindner et al, 
Phys. Rev. Lett. {\bf 95} 040401 (2005). 

\bibitem{photo-double-1} K. Kreidi et al, 
Science {\bf 318} 949 (2007).  

\bibitem{photo-high-energy} J. Fernandez et al, 
Phys. Rev. Lett. {\bf 98} 043005 (2007).  

\bibitem{Ako07} D.~Akoury \textit{et al.}, 
Science 318, 949 (2007).

\bibitem{Zim08} B.~Zimmermann \textit{et al.}, 
Nature Physics 4, 649 (2008).

\bibitem{photo-double-2} K. Kreidi et al, 
Phys. Rev. Lett. {\bf 100} 133005 (2008).  

\bibitem{lin} C. D. Lin et al, 
J. Phys. {\bf B 43}, 122001 (2010).     

\bibitem{cherepkov} 
N. A. Cherepkov et al, Phys. Rev. {\bf A 82}, 
023420 (2010).  


\bibitem{e-2e} D. S. Milne-Brownlie et al, 
Phys. Rev. Lett. {\bf 96} 233201 (2006).  

\bibitem{Sto01} N.~Stolterfoht \textit{et al.}, 
                Phys. Rev. Lett. 87, 023201 (2001).

\bibitem{Mis04} D.~Misra \textit{et al.},
                Phys. Rev. Lett. 92, 153201 (2004).

\bibitem{cristina} C. Dimopoulou et al, 
Phys. Rev. Lett. {\bf 93} 123203 (2004).  

\bibitem{daniel} H. T. Schmidt et al, 
Phys. Rev. Lett. {\bf 101} 083201 (2008).  

\bibitem{doerner} L. Ph. H. Schmidt et al, 
Phys. Rev. Lett. {\bf 101} 173202 (2008).  

\bibitem{schulz} J.S. Alexander et al, 
Phys. Rev. {\bf A 78} 060701(R) (2008).  

\bibitem{double-capt-we} D. Misra et al,  
Phys. Rev. Lett. {\bf 102} 153201 (2009).  

\bibitem{we-2011} A. B. Voitkiv et al,  
Phys. Rev. Lett. 106, 233202 (2011). 

\bibitem{shaofeng-2014} S. F. Zhang et al, 
Phys. Rev. Lett. {\bf 112}, 023201 (2014).  

\bibitem{shaofeng-2018} Y. Gao, S. F. Zhang, 
X. L. Zhu, et al,  
Phys. Rev. {\bf A 97}, 020701(R) (2018). 

\bibitem{ref-on-beam-size}  
D. Fischer, A. B. Voitkiv, R. Moshammer, and J. Ullrich, 
Phys. Rev. {\bf A 68}, 032709 (2003). 

\bibitem{be+ecc} J. H. Macek and S. T. Manson, Chapter 53 in {\it Springer Handbook 
on Atomic, Molecular and Optical Physics.} (Springer, Berlin, 2006). 

\bibitem{cdw-eis} D. S. F. Crothers and J. McCann,   
J. Phys. {\bf B 16} 3229 (1983). 

\end{thebibliography}
\end{document}